# Magnon Valve Effect Between Two Magnetic Insulators

H. Wu,[1,2] L. Huang,[1,2] C. Fang,[1,2] B. S. Yang,[1,2] C. H. Wan,[1,2] G. Q. Yu,[1,2] J. F. Feng,[1,2] H. X. Wei,[1,2] and X. F. Han[1,2,*]

[1]*Beijing National Laboratory for Condensed Matter Physics, Institute of Physics, Chinese Academy of Sciences, Beijing 100190, China*

[2]*University of Chinese Academy of Sciences, Beijing 100049, China*

*Corresponding author. E-mail：xfhan@iphy.ac.cn (X. F. Han)

**Abstract:**

The key physics of the spin valve involves spin-polarized conduction electrons propagating between two magnetic layers such that the device conductance is controlled by the relative magnetization orientation of two magnetic layers. Here, we report the effect of a magnon valve which is made of two ferromagnetic insulators (YIG) separated by a nonmagnetic spacer layer (Au). When a thermal gradient is applied perpendicular to the layers, the inverse spin Hall voltage output detected by a Pt bar placed on top of the magnon valve depends on the relative orientation of the magnetization of two YIG layers, indicating the magnon current induced by spin Seebeck effect at one layer affects the magnon current in the other layer separated by Au. We interpret the magnon valve effect by the angular momentum conversion and propagation between magnons in two YIG layers and conduction electrons in the Au layer. The temperature dependence of magnon valve ratio shows approximately a power law, supporting the above magnon-electron spin conversion mechanism. This work opens a new class of valve structures beyond the conventional spin valves.

**Main Text:**

A spin valve, which comprises of a nonmagnetic metallic [1,2] or insulating layer [3,4] sandwiched by two metallic ferromagnetic layers, has been widely adopted as an essential building in spintronic devices such as magnetic reading heads of hard disk drives (HDD) [5], magnetic random-access memory (MRAM) [6,7], and spin logic [8,9], etc. Information of a spin valve is encoded through relative magnetization directions of two ferromagnetic layers. Although there are a variety of electric means to write a spin valve including the spin-transfer torques [10,11] and spin-orbit torque [12,13], the reading of the spin valve is exclusively based on giant magnetoresistance or tunnel magnetoresistance in which spin-polarized electrons propagate from one magnetic layer to the other. Thus, a necessary condition for a functional spin valve is that the magnetic layers must be metallic and possess a large electron spin polarization.

Fundamentally, the spin can be carried without electrons such as magnons, photons, neutrons, and so on, among them magnons have attracted a great deal of interests recently [14-17]. Magnons are quasi-particles of spin wave, which represent the coherent collective excitation in magnetic systems, and each quantized magnon carries a spin angular momentum of $-\hbar$. The wave property of magnons provides some unique features that are unavailable in electron based spintronic devices: firstly, magnons provide the long-distance spin information propagation without Joule heating, which could drastically reduce the power consumption of spintronic devices. Secondly, the modulation of phase parameter provides another degree of freedom to information processing, which could realize the non-Boolean logic operation. Moreover, the quantum property of magnons inspires some macroscopic quantum phenomena such as spin superfluid [18], magnon Josephson effect [19], and so on.

The concept of magnonics was proposed to study the magnon based fundamental physics and potential applications very recently [14]. As we know, transistors and spin valves act as the basic unit of semiconductor and spintronic devices respectively. Therefore, in magnonic devices and circuits, we need a basic building block - magnon valve to accomplish functional information processing and data storage. The typical magnon valve structure consists of two ferromagnetic layers that are separated by a space layer, and the magnon valve effect means that the magnon current transmission coefficient could be controlled by the relative orientations of two ferromagnetic layers. In the ideal case, all the magnon current could pass through the magnon valve at the parallel magnetization state and be blocked at the antiparallel magnetization state. Especially, ferromagnetic insulator (FMI) based magnon valve is a promising candidate, because the insulating property prohibits any electron motion, and magnons become the sole spin information carriers in FMI.

Recent studies on spin Seebeck effect (SSE) [20-25] and magnon drag effect [26-29] have demonstrated that magnon current in ferromagnetic insulators can be generated by either thermal gradient or electron spin injection. More importantly, the magnon current can convert into an electron spin current at the interface between the FMI and the nonmagnetic metal (NM) [26-29]. And also, some theoretical works also investigated the magnon-mediated pure spin current transport and spin transfer torque between two FMI layers [30-33]. These studies lay the foundation of the experimental investigation of magnon valve structures in which the metallic magnetic layers are replaced by ferromagnetic insulator layers.

In this work, we propose to experimentally investigate the spin transport in the magnon valve structure FMI/NM/FMI. When a thermal gradient is applied to the magnon valve, the magnon current in one FMI layer would be affected by the magnon current in the other FMI layer, mediated through the electron spin current in the NM layer. If one measures the magnon current across the magnon valve by depositing a heavy metal Pt layer, one would find that the inverse spin Hall effect (ISHE) [34,35] voltage depends on the relative orientation of the magnetization of the two FMI layers, i.e., the magnon valve effect, see Fig. 1(a).

We choose yttrium-iron-garnet (YIG) as both top and bottom FMI layers. YIG is known for its low Gilbert damping factor ($\alpha \sim 10^{-4}$) [36] and wide band gap ($E_g = 2.85$ eV) [37,38]. Magnon valve structures YIG(40)/Au($t_{Au}$)/YIG(20)/Pt(10) (thickness in nanometers, from bottom to top) were deposited on 300-μm $Gd_3Ga_5O_{12}$ (GGG) (111) substrates by an ultrahigh vacuum magnetron sputtering system (ULVAC MPS-4000-HC7 Model), and the base pressure of the sputtering chamber was $1 \times 10^{-6}$ Pa. After deposition, an 800 °C annealing in the air was carried out to improve the crystal structure of YIG layers [39]. The multilayers were then patterned into the 100 μm × 1000 μm stripe by standard photolithography technique combined with Ar-ion etching, and then a 200 nm MgO insulating layer and a 10 nm Au heating electrode (100 μm×1000 μm) were fabricated on top of the magnon valve to enable the longitudinal temperature gradient $\nabla T$ via on-chip Joule heating. The cross-sectional scanning and high-resolution transmission electron microscopy (STEM and HRTEM) results of the YIG(40)/Au(15)/YIG(20)/Pt(10 nm) magnon valve structure were measured by a Tecnai G2 F20 S-TWIN system. The magnetic field dependence of the magnetization was measured by a vibrating sample magnetometer (VSM EZ-9, MicroSense). All magnetotransport measurements were performed in a physical property measurement system (PPMS-9T, Quantum Design) with a horizontal rotator option.

In YIG(40)/Au(15)/YIG(20)/Pt(10 nm) magnon valve structure, the well-defined epitaxial single crystal structure of YIG was formed on the GGG (111) surface, and the selected area electron diffraction (SAED) patterns show that the YIG film was grown along the (111) direction, as shown in Fig. 1(b). From Fig. 1(c), both bottom YIG/Au and top Au/YIG interfaces are atomically sharp, which promises a reduced diffusive scattering and a higher rate of conversion between magnon current and electron spin current at the interfaces. However, one would expect the crystal quality of

YIG deposited on Au is worse than that on GGG. Such difference is reflected in the coercive fields of the bottom and top YIG layers: the bottom YIG layer has a coercivity of 0.7 Oe while the top YIG layer has more than one order of magnitude higher coercivity (47 Oe). The separation of the coercivity of two YIG layers is necessary for generating an anti-parallel configuration of the magnon valve structure as long as the coupling field from either magnetostatic coupling or indirect Ruderman–Kittel–Kasuya–Yosida (RKKY) exchange interaction is sufficiently weak. We have changed the thickness of the Au layer $t_{Au}$ from 2 nm to 15 nm, and find that the coupling is weak enough for a well-separated magnetization reversal of bottom and top YIG layers when $t_{Au}$ exceeds 6 nm (see in Supplemental Materials).

The temperature gradient $\nabla T$ in YIG(40)/Au(15)/YIG(20)/Pt(10 nm) magnon valve structure is created by applying a 20 mA electric current in the heating electrode on the top of the magnon valve with a thick MgO spacer layer in between. The temperature gradient would exist for both top and bottom YIG layers, generating a local magnon current. Since the Pt layer is in contact with the top YIG layer, the ISHE voltage measured in Pt would be proportional to the total magnon current of the top YIG layer. Aside from the magnon current associated with the local temperature gradient in the top YIG layer, the magnon current in the bottom YIG layer can flow into the top YIG layer by first converting into an electron spin current in Au layer and subsequently converting back to the magnon current in the top YIG layer. Thus, depending on whether the magnetization of the two YIG layers in the parallel or anti-parallel state, the total magnon current would be sum or difference of these two magnon currents. Figs. 2(a) and 2(b) show both the hysteresis loop (*M-H* curve) and the ISHE voltage-magnetic field loop ($V_{ISHE}$-*H* curve) respectively. The *M-H* loop illustrates a clear two-step magnetization reversal. Since the magnetic moment of the bottom YIG is made to be 2 times that of the top YIG, the relatively sharp and large magnetization jump at the smaller field indicates the magnetization reversal of the bottom YIG layer. For the $V_{ISHE}$-*H* loop, a clear difference is seen for the magnetization of two YIG layers in parallel and anti-parallel states. If we defined a magnon valve ratio $MVR = (V_{\uparrow\uparrow} - V_{\downarrow\uparrow}) / (V_{\uparrow\uparrow} + V_{\downarrow\uparrow})$, where $V_{\uparrow\uparrow}$ ($V_{\downarrow\uparrow}$) is the measured ISHE voltage in Pt for the two YIG layers in the parallel (anti-parallel) state, we found, for example, *MVR* is 11 % for a 15 nm Au interlayer.

Next, a series of structures were designed to rule out possible artifacts of our observed effect.

When the interlayer Au in YIG(40)/Au(15)/YIG(20)/Pt(10 nm) structure [Fig. 3(a)] was replaced by a 10 nm-thick insulating MgO layer, the magnon current from the bottom YIG is completely blocked and thus one would not expect any magnon current contribution from the bottom layer. Indeed, the $V_{ISHE}$-$H$ loop, as seen in Fig. 3(b), only follows the magnetization of the top YIG layer, and the magnetization reversal of the bottom layer does not affect ISHE voltage. On the other hand, the sample with the bottom layer only, YIG(40)/Pt(10 nm), display a sharp transition within ±10 Oe, as shown in Fig. 3(c), indicating a normal spin Seebeck signal for the bottom YIG layer. While for the Au(15)/YIG(20)/Pt(10 nm) sample, the inserted Au layer between GGG and YIG leads to a larger coercive filed with non-square hysteresis, as shown in Fig. 3(d). These controlled experiments support our proposed magnon valve effect: the ISHE voltage depends on the relative orientation of the magnetization of the two YIG layers.

Another test of the magnon valve effect is to study the interlayer Au thickness dependence. When the thickness of Au exceeds its spin diffusion length (SDL) [40,41], the magnon current in the bottom layer is unable to reach the top layer and thus the magnon valve ratio diminishes. In Fig. 4(a), we show $MVR - t_{Au}$ relation for $t_{Au}$ from 6 nm to 15 nm. If the curve is fitted by a simple exponential decay function, we can obtain the spin diffusion length of Au (300 K) to be 15.1 nm, which is close to the 12.6 nm from spin pumping measurement [42].

Since the magnon valve effect comes from the magnon current propagating from the bottom to top YIG layers, mediated by the electron spin current of the spacer layer, the study of the temperature dependence would reveal the electron-magnon spin conversion efficiency [26-29,43] at the two interfaces. Fig. 4(b) shows the temperature dependence of the magnon valve ratio. As expected, the magnon valve ratio decreases as the temperature drops; this is consistent with the magnon transport in which the number of magnon carriers is fewer at lower temperature.

A simple model can be used to quantitatively estimate the observed temperature dependence. The total magnon current in the top YIG layer comes from the local temperature gradient as well as the magnon flow from the bottom layer to the top layer. In fact, the magnon current generated by SSE would increase with increasing the thickness of YIG, $\rho = \dfrac{\cosh(t_{FM}/l_m) - 1}{\sinh(t_{FM}/l_m)}$, where $\rho$ is a factor that represents the effect of the finite YIG layer thickness, $t_{FM}$ is the thickness of YIG, and $l_m$ is the magnon diffusion length [44]. After considering the thickness dependence of SSE and using the 70 nm $l_m$ in previous work [44], the magnon current from bottom YIG layer (40 nm) is 1.96 times of that from top YIG layer (20 nm). The magnon current from the bottom layer would suffer three reduction factors to reach the top layer: the magnon-to-electron spin conversion rate at the bottom YIG/Au interface $G_{me}$, the electron-to-magnon spin conversion rate at the top Au/YIG interface $G_{em}$ and spin current loss in the Au layer. Since the ISHE voltage is proportional to the total magnon current in the top YIG layer, we could write $V_{\uparrow\uparrow(\downarrow\uparrow)} = a\nabla T \left[ 1 \pm 1.96 G_{me} G_{em} e^{\left(-\frac{d}{\lambda}\right)} \right]$, where $a$ is the spin Seebeck coefficient, $\lambda$ and $d$ are the spin diffusion length and the thickness of the Au layer, and we neglect the magnon current decay in YIG layers in this formula. Thus the magnon valve ratio is $MVR = \frac{V_{\uparrow\uparrow} - V_{\downarrow\uparrow}}{V_{\uparrow\uparrow} + V_{\downarrow\uparrow}} = 1.96 G_{me} G_{em} e^{-\frac{d}{\lambda}}$. The spin conversion rates had been previously calculated [29], $G_{me} G_{em} \propto T^{\frac{5}{2}}$, apart from the offset signal, indeed the experimental data fits the $T^{5/2}$ temperature dependence very well, as seen in Fig. 4(b), which is consistent with the spin conversion theory. The offset signal is partly due to the on-chip heating that makes the temperature of the sample is significantly higher (about 24 K) than the temperature of the control. Further study is needed to map out the temperature dependence of the magnon valve ratio.

When a magnetic field of 5 kOe rotates in the sample plane such that the magnetizations of two YIG layers are parallel to the magnetic field, the ISHE voltage displays a perfect sine angular dependent, $V_{ISHE} \propto \sin \alpha$, as shown in Fig. 4(c), in agreement with the conventional spin Seebeck behavior [20-25]. The amplitude of the sine relation is proportional to the square of the heating current $I$, as shown in Fig. 4(d), indicating the temperature gradient $\nabla T$ created by the on-chip heating scales as the heating power, as expected.

In conclusion, we have fabricated the YIG/Au/YIG/Pt magnon valve structure and investigated the thermally driven magnon current transport across the multilayers. The observed large magnon valve ratio supports the notion that the magnon current transmission between two magnetic insulating layers mediated by a nonmagnetic metal has high efficiency. Magnon valve ratio can be further improved via improving the spin conversion efficiency at FMI/NM interface and optimizing the materials and thickness of FMI and space layers. Magnon valves could be used to manipulate the transmission coefficient of magnon current, which have potential applications in magnon based logic, memory and on-off switching devices. Utilizing magnetic insulators rather than magnetic metals for spintronic devices has superior advantages in terms of low energy consumption, and the present results open a door for fundamental research and device application beyond those based on conventional spin valve structures.

**Acknowledgements:**

We gratefully thank S. Zhang for enlightening discussions and theoretical help. This work was supported by the National Key Research and Development Program of China [Grant Nos. 2016YFA0300802 and 2017YFA0206200], the National Natural Science Foundation of China (NSFC) [Grant Nos. 11434014, 11404382, and 51620105004], the Strategic Priority Research Program (B) of the Chinese Academy of Sciences (CAS) [Grant No. XDB07030200].

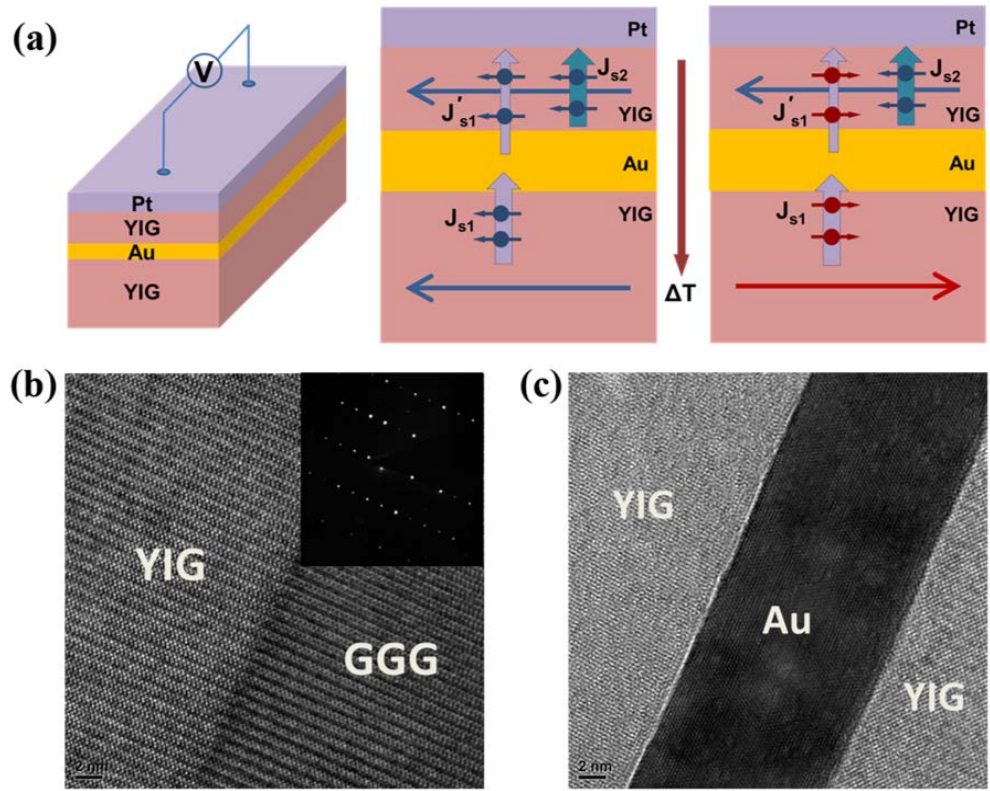

FIG. 1. (a) Illustration of the magnon valve effect: when a temperature gradient is applied, the magnon current in the top YIG comes from two sources. One is generated by the presence of the local temperature gradient and the other is the magnon current injected from the bottom YIG layer. If the magnetization directions of the YIG layers are parallel (anti-parallel), these two magnon currents are additive (subtractive). Since the inverse spin Hall (ISHE) voltage measured by the Pt layer is proportional to the total magnon current through the top layer, a magnon valve effect is observed. (b) The cross-sectional scanning transmission electron microscopy (STEM) and selected area electron diffraction (SAED) patterns of GGG/YIG interface. (c) The cross-sectional high-resolution transmission electron microscopy (HRTEM) of YIG/Au/YIG region. The magnon valve structure measured in (b) and (c) is GGG/YIG(40)/Au(15)/YIG(20)/Pt(10 nm).

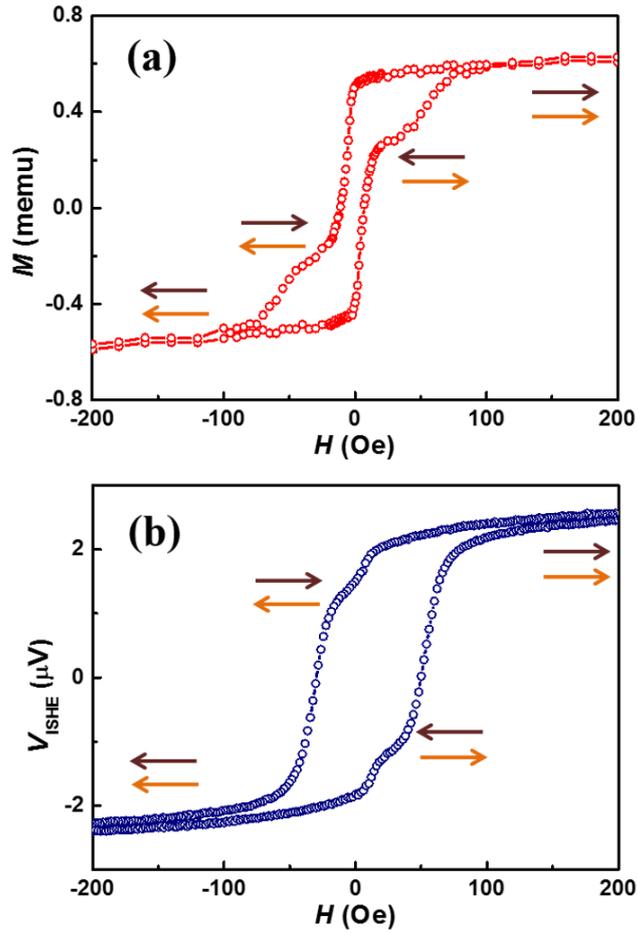

FIG. 2. The magnetic and magnon transport properties of the magnon valve. (a) Magnetization of the magnon valve structure GGG/YIG(40)/Au(15)/YIG(20)/Pt(10 nm) as a function of the magnetic field applied in the plane of the layers. The arrows indicate the magnetization directions of the two YIG layers. (b) The ISHE voltage in Pt as a function of the magnetic field for the same magnon valve structure in the presence of the temperature gradient created by a 20 mA electric current applied at the heating electrode.

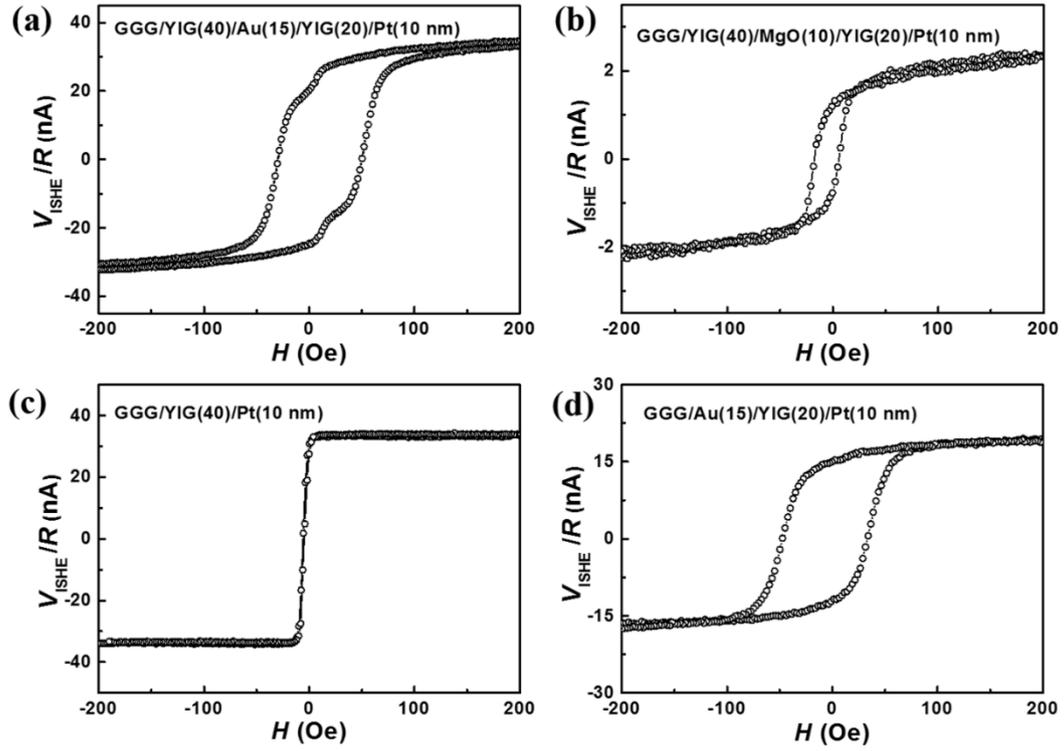

FIG. 3. The ISHE voltage of different samples for controlled study. (a)-(d) Sample structures are marked in the figures, and the ISHE voltage is normalized by the resistance of the Pt detector to eliminate the sample-to-sample variation.

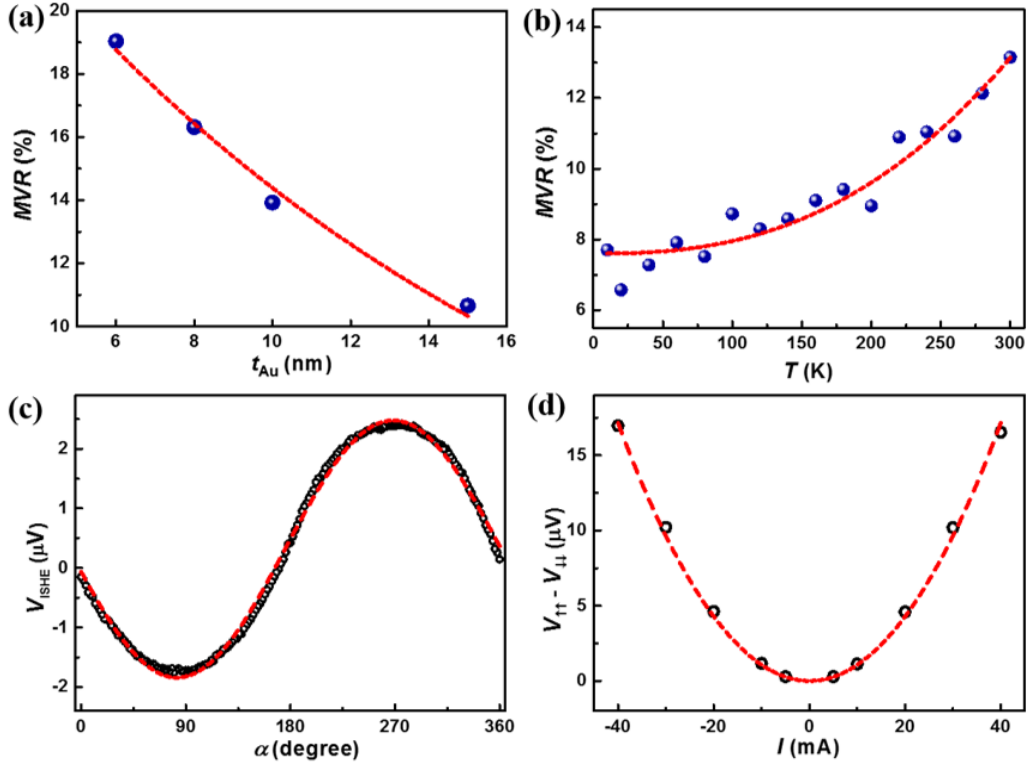

FIG. 4. Thickness, temperature, magnetization direction, and heating current dependences of the magnon valve effect. (a)-(d) were measured in GGG/YIG(40)/Au(15)/YIG(20)/Pt(10 nm) sample. (a) The interlayer Au thickness dependence of magnon valve ratio *MVR*. The dashed line shows the exponential decay fitting curve. (b) The temperature dependence of *MVR*, and the dashed line shows the $T^{5/2}$ fitting curve. (c) The ISHE voltage as a function of the angle between the directions of the voltage probe and the in-plane magnetic field. The magnetic field (5 kOe) is large enough such that both YIG layers are in parallel with the magnetic field. The dashed line shows the sine fitting curve. (d) The heating current dependence of $V_{\uparrow\uparrow} - V_{\downarrow\downarrow}$. The dashed line shows the parabolic fitting curve.